\documentclass[%
 reprint,
 amsmath,amssymb,
 aps,
]{revtex4-1}

\usepackage{graphicx}
\usepackage{dcolumn}
\usepackage{bm}

\begin{document}

\preprint{APS/123-QED}

\title{Modeling influenza-like illnesses through composite compartmental models}

\author{Nir Levy}
\affiliation{Microsoft Israel, Herzeliya 46733, Israel}%
 \email{nirlevy@microsoft.com}
\author{Michael Iv}%
\affiliation{Microsoft Israel, Herzeliya 46733, Israel}%
\author{Elad Yom-Tov}%
\affiliation{Microsoft Israel, Herzeliya 46733, Israel}%

\date{\today}

\begin{abstract}

Epidemiological models for the spread of pathogens in a population are usually only able to describe a single pathogen. This makes their application unrealistic in cases where multiple pathogens with similar symptoms are spreading concurrently within the same population. Here we describe a method which makes possible the application of multiple single-strain models under minimal conditions. As such, our method provides a bridge between theoretical models of epidemiology and data-driven approaches for modeling of influenza and other similar viruses.

Our model extends the Susceptible-Infected-Recovered model to higher dimensions, allowing the modeling of a population infected by multiple viruses. We further provide a method, based on an overcomplete dictionary of feasible realizations of SIR solutions, to blindly partition the time series representing the number of infected people in a population into individual components, each representing the effect of a single pathogen. 

We demonstrate the applicability of our proposed method on five years of seasonal influenza-like illness (ILI) rates, estimated from Twitter data. We demonstrate that our method describes, on average, 44\% of the variance in the ILI time series. The individual infectious components derived from our model are matched to known viral profiles in the populations, which we demonstrate matches that of independently collected epidemiological data. We further show that the basic reproductive numbers ($R0$) of the matched components are in range known for these pathogens.

Our results suggest that the proposed method can be applied to other pathogens and geographies, providing a simple method for estimating the parameters of epidemics in a population.

\end{abstract}

\pacs{Valid PACS appear here}
\maketitle


\section{Introduction}


Compartmental models, first suggested by Hamer \cite{Hamer1906} and later developed by Kermack and McKendrick \cite{McKendrick1927} describe the spread of an infection through the interaction between the parameters of a pathogen and three populations: susceptible individuals, infective individuals and recovered individuals. These models are commonly known as susceptible-infected-recovered (SIR) models. SIR models were originally developed to model individual pathogens. More recently, these models were extended to multiple compartments, which allows the modeling the dynamics of multiple disease strains \cite{Andreasen1997JMB,Gog2002PNAS,Gupta1998Nature,Crepey2015ISIRV} and cross-immunity in an age structured model \cite{castillo1989,Gog2002JMBoscillatoryDynamics,Gog2002JMB} . 

A different approach to modeling disease spread is a data-driven approach, using the most recent known disease load and\\or proxy data to disease load. The former is usually highly accurate but is gathered with some delay and at lower temporal and spatial resolution, compared to the latter, which is less accurate but can be collected in near real time with relatively high spatial resolution. Examples of such proxy data have, in recent times, focused on Internet data such as search queries \cite{polgreen2008,Brilliant2009NatureLett,Zhang2017ScientificReports,Lipsitch2013NatureCommunications}, social media postings \cite{lampos2015,yom2015estimating} or other Internet data \cite{mciver2014}. Social networks often reflect dynamics of aggregation of types and subtypes of influenza, and commonly serve as crude estimation to the real number of infected individuals. Monitoring influenza by type and subtype not only provides more detailed observational content but supports more accurate forecasting \cite{Shaman2017AJE}.



However, when tracking disease with such proxy data it is usually impossible to distinguish between similar diseases because data are usually related to symptoms of the disease. In the case of influenza, these include low specificity symptoms such as cough, sore throat, and fever of $ \geq 37.8 ^{\circ} C $. Hence, tracking of influenza is usually replaced by tracking of influenza-like illnesses (ILI). ILI includes diseases with similar symptoms, such as respiratory syncytial virus (RSV) and parainfluenza. If (as is usually the case) several diseases overlap in time and space, ILI rates cannot be modeled by simple SIR models, and the important link between data-driven disease tracking and tracking via epidemiological models is lost. 

Prior research has attempted to bridge this gap. However, simple application of multiple SIR models is limited by the dimensionality of the resulting problem. Modeling of $n$ viral strains in a simple history-based model requires the solution to $O(2^n)$ equations, making it difficult to include more than a few strains. As a result, most previous work using a population level model has either focused on a small number of strains \cite{Sasaki2009TPB,Andreasen1997JMB,Gomes2002PRSB}, or reduced the dimensionality of the model by making certain strategic assumptions \cite{gog2002dynamics,Kryazhimskiy2007}. In particular, it is possible to reduce the history-based equation system to $O(n)$ equations if all strains are tracked, but the order in which they are seen is not \cite{Ferguson2002MAERID,Minayev2008JRSI}. This limits the number of potential strains that can be studied unless computationally intensive individual-based models are used \cite{Ferguson2003NatureLett,Tria2005SM}.


Thus, here we focus on modeling multiple pathogens using an ensemble of SIR models. We show that the temporal dynamics of ILI can be blindly partitioned into multiple SIR models in a data-driven manner and show that the resulting models correspond to known pathogens.

\section{Modeling}

Our modeling approach is based on the basic SIR model, due to Kermack-McKendrick \cite{Kermack700}. It
describes the evolution of an infectious agent in a population using a system of ordinary differential equations:

\begin{equation}
	\frac{dS}{dt} = - \beta S I , \quad
    \frac{dI}{dt} = \beta S I - \gamma I, \quad
    \frac{dR}{dt} = \gamma I
    \label{scalar_sir}
\end{equation}

where $\beta > 0$ is the infection rate and $\gamma > 0$ is the recovery rate. To improve this model and account for multiple viruses infecting the same population we suggest to transform (\ref{scalar_sir}) to be multidimensional, thus representing the dynamics of a population infected by multiple viruses and virus strains. To this end, we assume a $v$ dimensional space, where $v$ denotes the number of distinct viruses existing in a specific population prone to infection. In this is the case, $S$, $I$, and $R$ take the form of square matrices with dimension $v$. The specific elements in these matrices, e.g. $S_{ij}$, are defined for viruses of type $i$ in case that $i=j$, or when $i{\neq}j$, as a mutation of virus $i$ to virus $j$. We force $0$ to specific elements in $S$, $I$, and $R$ when the option to mutate does not exist or its likelihood is negligible. $\beta$ in its matrix form accounts for the infection rate and the ability to cross infect. Similarly, $\gamma$ is defined such that in the case of $i{\neq}j$ we account for a different recovery period for people who are infected by a mutation and were infected in the past by a similar virus.

In the general case the new dynamics can be defined as:

\begin{equation}
	\begin{gathered}
    \begin{aligned}
		\frac{d \mathbf{S}}{dt} &= - \mathbf{I \bm\beta^\intercal S} \\
  	  	\frac{d \mathbf{I}}{dt} &= \mathbf{S \bm\beta^\intercal I - \bm\gamma^\intercal I} \\
  	  	\frac{d \mathbf{R}}{dt} &= \mathbf{\bm\gamma^\intercal I} \\
 	  	\label{matrix_sir}
	\end{aligned}
	\end{gathered}
\end{equation}

representing a pool of $v$ viruses effecting a population of size $N=constant$. This approach enables us to capture the composite seasonal dynamics, as will be shown in the following sections. The state of the system at time $t$ is defined by the two matrices $\mathbf{S}(t)$ and $\mathbf{I}(t)$. In the simplest case these matrices, as well as $\bm\beta$ and $\bm\gamma$ will take the diagonal form, allowing no mutations or cross infections.

Degrading the dynamics to allow only diagonal states represents a simple dynamic in which the different viruses affect people independently. That is, the dynamic will follow (\ref{scalar_sir}) and $N$ is preserved for each virus type independently. This degradation assumes no interaction between different virus types and no increase in, for example, the likelihood of infection in people infected or recovered from other viruses. 

In a degraded system as described above, the $L_1$ norm (the sum of the diagonal elements) of the state matrix $I(t)$ is equivalent to the total size of the infected sub-population at a given time $t$. If influenza-like illnesses are modeled using Equations (\ref{matrix_sir}), then $L_1$ represents the ILI rate. Moreover, since each virus is independent of other viruses, its equations can be solved independently of others. We take this approach in the next sections, and show how this model can be fit to real seasonal data.

\section{Methods}

\subsection{Data collection}

\subsubsection{Twitter data}

We collected all messages from the Twitter social network, also known as tweets, originating from England during five consecutive influenza seasons, from 2012 to 2017. Each year we collected these data from October 1st to April 30th of the following year. Tweets were identified as originating in England if they had GPS coordinates embedded in them, and these coordinates were within England. The total number of tweets per season is shown in Table \ref{tbl:data}.

\begin{table}[th]
     \centering
    	\begin{tabular}{|l|c|}
        \hline
        \textbf{Season} & \textbf{Number of ILI tweets} \\
        \hline
        2012-2013 & 133,269 \\
        2013-2014 & 187,289 \\
        2014-2015 & 91,552\\
        2015-2016 & 3,493 \\
        2016-2017 & 2,208 \\
        \hline
        \hline
    \end{tabular}
    \caption{Number of ILI tweets per season}
    \label{tbl:data}
\end{table}

Following the methodology reported in Yom-Tov et al. \cite{yomtov2015} (where full details of the methodology are provided), we identified  Twitter messages that were likely related to ILI by constructing a large set of ILI-related terms and then narrowed them to contain only the most informative of these terms. We began by manually crafting a list of 217 textual markers related to or expressing symptoms of ILI. This list was narrowed down by constructing a linear prediction model to obtain the best correlation between the ILI rates gathered by the Royal College of General Practitioners (RCGP) and published by Public Health England (PHE) and the number of times each term was mentioned in tweets during the same time period. We then selected the 20 phrases which had the largest weight in the model, and retained only those tweets which contained one or more of these terms. The terms are listed in Table \ref{tbl:terms}. 

The number of people who mentioned on of these 20 terms each day is our estimated ILI rate, as computed from Twitter data.

\begin{table}[tb]
\begin{center}
\begin{tabular}{lll}
  \hline
  Term & & \\
	\hline
bad cough &
bed flu &
chest infection \\
chesty cough &
cold flu &
cough \\
cough syrup &
coughing &
feel sick \\
flu &
food feel sick &
headache night \\
illness &
man flu &
shivering \\
throat cough &
vomit &
vomiting \\
waking headache &
worst cough & \\
\end{tabular}
\end{center}
\caption{Terms used for the first stage of tweet filtering}
\label{tbl:terms}
\end{table}

\subsubsection{Public Health England (PHE) data}

We extracted the weekly ILI rate reported from RCPG in the PHE Weekly National Influenza Reports. Additionally, we extracted the fraction of samples which tested positive for influenza, as reported by the Respiratory DataMart System. These included the fractions for each of the following strains: Influenza A(H3), A(H1N1)pdm09, A(not subtyped) and B. We approximate the fraction of each strain in the population as the product of the ILI rate by the fraction of detections. Also recorded were the positivity for RSV, Rhinovirus, Parainfluenza, Adenovirus, and human metapneumovirus (hMPV). Thus, weekly data for 9 viral strains was available to us as ground truth data. 
 
\subsection{Blind decomposition of aggregate time series}

Matching pursuit (MP) \cite{mallat1993} is a method for decomposing a signal into a linear combination of waveforms drawn from a redundant dictionary of functions. Here we apply MP with a dictionary of functions which are instantiations of the number of infected individuals over time, in solutions to SIR models with different parameters.

In a simple model of a compartmental SIR,
there are four parameters, shown in Table \ref{tbl:dictionaryParameters}. Additionally, for the case of MP, a phase parameter, $\theta$, is added to allow infections that begin later in a season. Finally, each function is multiplied by a gain parameter.

In the proposed solution, the dictionary of functions is comprised of the number of infected individuals over time for each valid solution of the SIR equations in a 10-point grid with the parameter values shown in Table \ref{tbl:dictionaryParameters}. Additionally, in order to remove solutions which were unlikely to be representative of actual seasonal viruses, we required solutions to have 50\% or more of infected individuals during 15\% or more of the days in the season. That is, solutions which showed most individuals being infected in a very short time (less than 15\% of the season) were removed as unfeasible.   

\begin{table*}[th]
     \centering
    	\begin{tabular}{|l|c|c|}
        \hline
        \textbf{Parameter} & \textbf{Range} & \textbf{Spacing} \\
	    \hline
        Size of the population ($N$) & $(10^5, 10^8)$ & Logarithmic \\
        The number of individuals infected at time zero ($I(0)$) & $(10^1, 10^3)$ &  Logarithmic \\
        Basic reproduction number ($R_0$) & $(0.7, 5)$ & Linear \\
        Rate of recovery ($\gamma$) & $10^{-6}, 10^{-2}$ & Logarithmic \\
        Phase ($\theta$) & $0, 100$ & Linear \\
        \hline
        \hline
    \end{tabular}
    \caption{Number of ILI tweets per season}
    \label{tbl:dictionaryParameters}
\end{table*}

MP finds $n$ functions from the dictionary which best match the ILI time series through a greedy approximation process. In our analysis the squared error was used as the criteria for matching. 
The number of functions was selected such that adding another component did not increase the value of the model fit ($R^2$) by more than 1\%. 
Since $N$ is assumed to be identical for all infectious agents, MP is run for each value of $N$, and the population size which reached the lowest squared error is taken as the best solution. 

\subsection{Matching individual components to known infections}

The result of MP is a set of functions from the dictionary, each of which has a corresponding set of SIR parameters. We match each of these functions to the time series of tracked diseases reported by PHE by finding the time series of the virus which has the highest Pearson correlation, without replacement.

\section*{Results}

\begin{figure}[t]
  \begin{center}
  \includegraphics[width=0.45\textwidth]{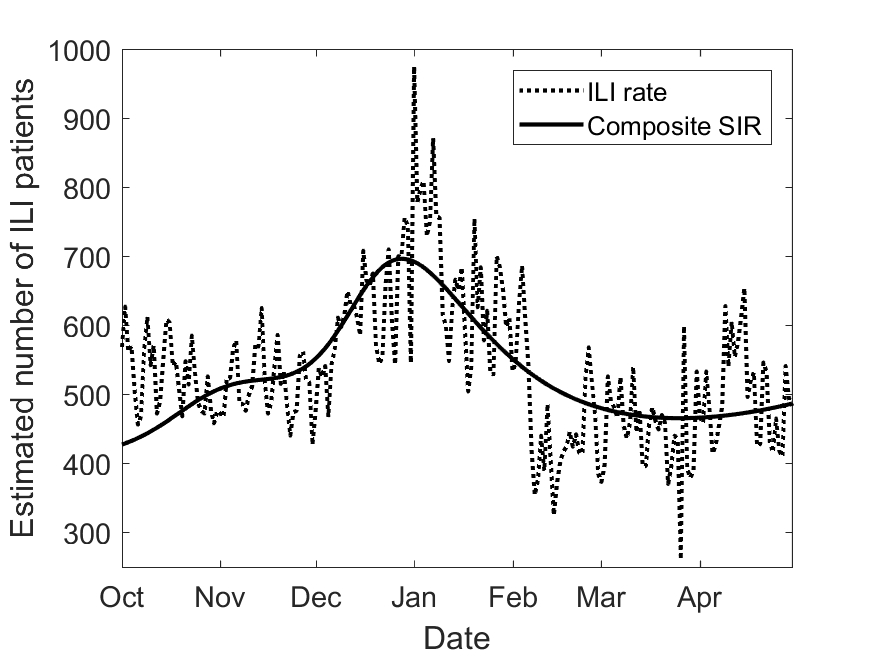}
\end{center}
\caption{The ILI rate inferred from Twitter postings and the ILI rate from the composite SIR model, for the 2012 season. The $R^2$ between the composite and actual ILI rate is 0.49.}
\label{fig:composite}
\end{figure}

Figure \ref{fig:composite} shows the ILI rate inferred from Twitter postings during the 2012 season, compared to the ILI rate from a composite of 5 SIR components. For this composite, $R^2=0.49$. During the 5 seasons examined, an average of 5 components was required so that adding components resulted in a lowering of $R^2$ by less than 1\%. The average $R^2$ for those seasons was 0.44. 

Table \ref{tbl:params} shows the average values of the SIR components matched to each virus, over the five seasons, for those viruses which were matched during at least 3 seasons. 

\begin{table}[th]
      \centering
     	\begin{tabular}{|l|c|c|}
         \hline
        \textbf{Virus} & \textbf{R0} & \textbf{I0}  \\
         \hline
 Influenza A(H3)	& 3.1 & 151 \\
  Influenza A(not subtyped) & 2.7 & 354 \\
 Influenza B				& 2.5 & 363 \\
 hMPV						& 3.7 & 297 \\ 
 Parainfluenza 				& 2.7 & 539 \\
 Rhinovirus 				& 1.2 & 604 \\
 RSV 						& 1.6 & 1000 \\
         \hline
         \hline
     \end{tabular}
     \caption{Average SIR component values for each virus across the five seasons.}
     \label{tbl:params}
 \end{table}

As noted above, PHE provides the fraction of reported influenza cases for each type, and the positivity rate for other viruses. After matching these reports over time with the SIR components (according to their Pearson correlation), we use linear regression to model the peak value for each virus during the season, from PHE reports, with the peak value of the SIR component multiplied by the gain parameter, as well as an indicator of whether this virus is reported in PHE through its detection rate (as for influenza) or its positivity in the population. 

The resulting model parameters are shown in Table \ref{tbl:peakfit}. As the Table shows, though the components were matched solely according to their shape over the season, there statistically significant correlation between the peak value of the SIR component computed from the Twitter ILI rate and the peak value of the matched virus as reported by PHE.  

\begin{table}[th]
     \centering
    	\begin{tabular}{|l|c|c|}
        \hline
        \textbf{Attribute} & \textbf{Slope (S.E.)} & \textbf{P-value}  \\
        \hline
Rate reported? 				& 13.0 (3.6) & 0.002 \\
Peak value of SIR component & 13.5 (6.1) & 0.037 \\
		\hline
        \hline
    \end{tabular}
    \caption{Model parameters for a linear estimator of the peak reported value in PHE for each virus at each season, as a function of the reporting type and the peak value of the matched SIR component. The $R^2$ for this model is 0.54 (n=24). }
    \label{tbl:peakfit}
\end{table}

\section*{Discussion}

Epidemiological models for the spread of pathogens in a population are usually only able to describe a single pathogen. This makes their application unrealistic in cases where multiple pathogens with similar symptoms are spreading concurrently within the same population. Here we describe a method which makes possible the application of multiple single-strain models under minimal conditions of independence. As such, our method provides a bridge between theoretical models of epidemiology and data-driven approaches for modeling of influenza and other similar viruses.

The proposed algorithm reaches an average $R^2$ of 0.44, which means that much of the daily variance in predicted ILI is explained by the model. This result was obtained by taking the naïve assumption of complete pathogen independency, where all state and parameter matrices take the diagonal form. In this case we assume that the ILI is represented by the $L_1$ norm of the $I(t)$ state matrix which represent simple counting of the infected population, regardless of the virus type. Note that by taking this assumption we allow each individual in the population to be infected by multiple viruses at a given point in time and thus $N$ is constant for every virus type. The non-diagonal elements in the model enable a more detailed analysis of the population, taking into account cross-pathogen dependencies. For example, it may be that early infection by one pathogen leads to lower immunity to or longer recovery from other pathogens. Thus, future work will focus on removing the independency assumption to evaluate if such removal can provide more explanatory models. We hypothesize that such improvement could be observed especially for mutations of the same virus, for viruses which are especially virulent, and in at-risk populations. The non-diagonal elements in $\bm\beta$ and $\bm\gamma$ can introduce these impacts to the dynamics of the  population state matrices.

We note that, after blind decomposition of the ILI time series, individual components could be matched to known viral profiles in the populations, and that their peak activity matches that of independently collected data (Table \ref{tbl:peakfit}). This matching suggests that the matched components are indeed related to the underlying pathogens. Another support to this finding is that the basic reproductive numbers ($R0$) of the matched components (Table \ref{tbl:params}) are in range known for these pathogens. Specifically, White et al. \cite{white2009} estimated $R0$ for the 2009 influenza A\H1N1 pandemic between 2.2 and 2.3, and a review of multiple studies reported a median $R0$ for seasonal influenza at 1.28 \cite{biggerstaff2014}. One estimate for RSV was found to be between 1.2 and 2.1 \cite{weber2001}. Thus, our estimates for $R0$ in influenza are relatively high compared to past studies, while the estimate for RSV is within known ranges. We hypothesize that the differences for influenza may be related to the differences in population demographics between past studies and ours, which was based on social media data.

\section{References}

\bibliography{influenza}

\end{document}